\documentclass[twocolumn]{aastex631}
\usepackage{subfigure}
\shorttitle{Observational Evidence of Particles Acceleration}
\shortauthors{Peng et al.}

\graphicspath{{./}{figures/}}
\usepackage{placeins}
\begin{document}

\title{Observational Evidence of Particles Acceleration by Relativistic Magnetic Reconnection in Gamma-ray Bursts}

\correspondingauthor{Rui-Jing Lu}
\email{luruijing@gxu.edu.cn}

\author{Cheng-Feng Peng} 

\author{Rui-Jing Lu}
\author{Wen-Qiang Liang}
\author{Zhe-geng Chen}

\affiliation{Laboratory for Relativistic Astrophysics, Physical Science and Technology College, Guangxi University, Nanning 530004, People’s Republic of China}

\begin{abstract}
Gamma-ray bursts (GRBs) as the most energetic explosions in the modern universe have been studied over half a century, but the physics of the particle acceleration and radiation responsible for their observed spectral behaviors are still not well understood. Based on the comprehensive analysis of the pulse properties in both bright GRB~160625B and GRB~160509A, for the first time, we identify evidences of particle acceleration by relativistic magnetic reconnection from the evolutionary behavior of the two spectral breaks ($E_{\rm p}$ and $E_{\rm cut}$). Meanwhile, the adiabatic cooling process of the emitting particles in the magnetic reconnection regions produces a relation between the spectral index and the flux. We also discuss the physics behind spectral energy correlations. Finally, we argue that the identification of an anticorrelation between $E_{\rm cut}$ and $L_{\rm iso}$ may opens a new avenue for diagnostics of the physics of the particle acceleration and radiation in a variety of astrophysical sources.

\end{abstract}



\section{Introduction} \label{sec:intro}
GRBs are one of the most energetic, and electromagnetically the brightest, transient phenomena in the Universe, and their light curves are very complex and the time scale of the variation is very short, thus the  inferred GRB source is very dense, which leads to a compactness problem \citep{Piran1999}. To overcome the compactness problem, their central engines, driven by either a black hole or a millisecond magnetar, must release an ultra-relativistically outflows (jets) \citep{Krolik1991,Fenimore1993,Woods1995,Baring1997}. However, over half a century from the discovery of GRBs, the energy composition of a relativistic jet, and even the dissipative and radiative mechanisms within such a jet remain open mysteries \citep{Kumar2015a}. A popular model for the energy dissipation invokes internal shocks in a matter-dominated fireball \citep{Goodman1986,Paczynski1986,Shemi1990,Paczynski1994, Rees1994}. Within such a picture, an initially hot fireball, composed of photons, electron–positron pairs, and a small amount of baryons, first converts most of its thermal energy into kinetic energy and then dissipates the kinetic energy in the internal (or sometimes external) shocks to power the observed GRB emissions. 

An alternative picture is magnetic energy dissipation within a Poynting-flux-dominated outflow \citep[e.g.][]{Usov1992,Thompson1994,Spruit2001,Drenkhahn2002,Lyutikov2003,
Giannios2008,ZhangYan2011}, which has been become more and more popular research topic, as it is believed that such a outflow can very effectively convert a significant fraction of its' magnetic energy into relativistic particles by collisionless magnetic reconnection, which could be made through first principles particle-in-cell (PIC) simulations \citep[e.g.][]{Zenitani2007, Cerutti2012,Guo2014, Guo2016, Werner2016, Granot2016, Beniamini2016, Sironi2016}. This simulations show that magnetic reconnection can accelerate particles into a power-law energy distribution with a high-energy exponential cutoff at $\gamma^{\prime}_{\rm cut}$ (the superscript prime $\prime$ denotes the quantities in the comoving frame hereafter). But the maximum energy attainable ($\gamma^{\prime}_{\rm cut}$) is dramatically different between two-dimensional (2D) \citep{Petropoulou2018,Hakobyan2021} and three-dimensional (3D) simulations \citep{Zhang2021,Werner2021}, i.e., in 3D particles can be fastly acceleraed with $\gamma^{\prime}_{\rm cut} \propto t^{\prime}$, in contrast, $\gamma^{\prime}_{\rm cut} \propto \sqrt{t^{\prime}}$ in 2D.    

Observationally, the typical GRBs usually show the Band spectrum \citep{Band93}, which typically peaks in $\nu F_{\rm \nu}$ between around 100 keV--1 MeV, sometimes, with an additional underlying power-law component extending to a higher energy end. This high-energy emission is believed to originate from a mechanism different from that for the low-energy component. If the spectral cutoff in the higher energy end is observed, one could either constrain the site of as a function of the bulk Lorentz factor \citep{GuptaZhang08}, or diagnose the jet composition, dissipative mechanisms and the physics of particle acceleration and radiation\citep{Zhang2009,Lefa2012}.  

Thanks to the wider energy range of Fermi/Gamma-ray Burst Monitor, the broadband spectrum for some bright GRBs can be studied from a few keV up to hundreds of GeVs. A spectral break around 0.4 GeV is detected for the first time in GRB~090926A \citep{Ackermann2011}. Following this way, more GRBs (i.e., GRB~100724B, GRB~160509A, GRB~160625B, and GRB~190114C) are reported to have a spectral cutoff at the highest-energy end \citep{Tang2015,Wang2017,Vianello2018,Chand2020,Ravasio2023}.

GRB~160625B was triggered by the Fermi/GBM on 2016 jun 25 at 22:40:16 UT ($T_0$) and its redshift ($z$) is 1.406 \citep{Xu2016}. It is one of the most powerful bursts with the isotropic energy $\sim E_{\rm iso}~3\times10^{54} erg$ \citep{Wang2017,Zhang2018}, and then has been a well-studied object and has several attractive properties: Its gamma-ray light curve has three distinct episodes separated by two quiescent times, which is explained through the transition from a fireball to Poynting flux-dominated jet \citep{Zhang2018}; The high-energy spectral cutoffs have been obviously identified in the first pulse in the second emission episode \citep{Wang2017, Lin2019}, at the same time, a significant linear optical polarization is also observed during the second emission episode \citep{Troja2017}. The facts clearly show that the outflow responsible for the bright gamma-ray emission is dominated by Poynting flux (also see \cite{Fraija2017}).  

Regardless of the dominant radiative processes responsible for the GRB production, it is expected that there should be a spectral cutoff ($E_{\rm cut}$) at the high energy end in the GRB spectrum due to the pair production \citep{Piran1999,Lithwick2001} and/or a high energy cutoff in the electron distribution. Once a spectral cutoff is observed, people \citep{Tang2015,Vianello2018,Lin2019,Chand2020,Li2023} usually attribute to the pair production and derive the bulk Lorentz factor $\Gamma$ of the outflow by assuming $\tau_{\gamma\gamma}(E_{\rm cut})=1$ \citep{Lithwick2001}. By applying the same method to GRB~160625B, \cite{Wang2017} found that the Lorentz factors derived from the cutoffs are well below the lower limits setted by the Compton scattering effect, which indicates that these cutoffs prefer trace the high energy cutoff in the electron distribution rather than being caused by the pair productions. If so, it's worth investigating what acceleration mechanisms are responble for the cutoff profile of electron energy distribution.

Interestingly, when performing time-resolved spectral analysis to the first pulse in the second emission of GRB~160625B, which is very smooth and well-shaped, \cite{Lin2019} found that, the Band function with the high-energy spectral cutoff ($E_{\rm cut}$)\footnote{Bandcut: https://asd.gsfc.nasa.gov/Takanori.Sakamoto/personal/} is best fit to the observed data, and $E_{\rm cut}$ extends to high energy with time during its decay phase (See in Fig.\ref{fig:160625B}), meaning that $E_{\rm cut}$ is negatively related to it's corresponding energy flux, different from the Amati relation \citep{Amati2002} and other spectral energy correlations, which have been found in the sub-MeV energy regime \citep[e.g.][]{Liang1996,Lloyd2000,Lu2010}. 

Similar spectral evolutionary feature like GRB~160625B is rare, but not unique. Extensive investigation into all bright GRBs with a smooth light curve pulse observed by Fermi satellite comes to another example of this in GRB~160509A, which was triggered by the Fermi/GBM on 2016-05-09 08:58:46.219 ($T_0$) in redshift 1.17 \citep{Tanvir2016}. It consists of soft precursor, following by two distinct main episodes, 0-40 s and 280-420 s \citep{Tam2017,Vianello2018}; It's first main episodes with the isotropic energy of $~ 1.47\times10^{52}$ erg \citep{Tam2017} exhibits a single-pulse profile with a smooth decay phase, in which the Bandcut can also provide a good description for all time-resoved spectra (see in Figure 2 of \cite{Vianello2018}. So, in this paper, we will investigate in more detail the underlying physics based on the two bright GRBs.

As mentioned above, both 2D and 3D PIC simulations show that magnetic reconnection can effectively accelerate particles into a power law followed by a high-energy exponential cutoff. So, we wander that, in the context of relativistic magnetic reconnection, whether or not PIC simulation can resemble the observations in the two bright GRBs, whose spectral evolutionary features is firstly shown in section \ref{sec:observations}. Then, in section \ref{sec:adiabatic}, we try to expain the orgin of the observations. Finally, in section \ref{sec:diss}, we present our conclusions and discussions. Throughout the paper we assume we assume the following cosmology: $\Omega_{\rm m} = 0.315$, $\Omega_{\Lambda} = 0.685$, and $H_0 = 67.4$\,km\,s$^{-1}$\,Mpc$^{-1}$ \citep{Planck20}.
   
\section{Observational features in both GRB~160625B and GRB~160509A} \label{sec:observations}

\begin{figure}
\centering
\includegraphics[scale=0.75]{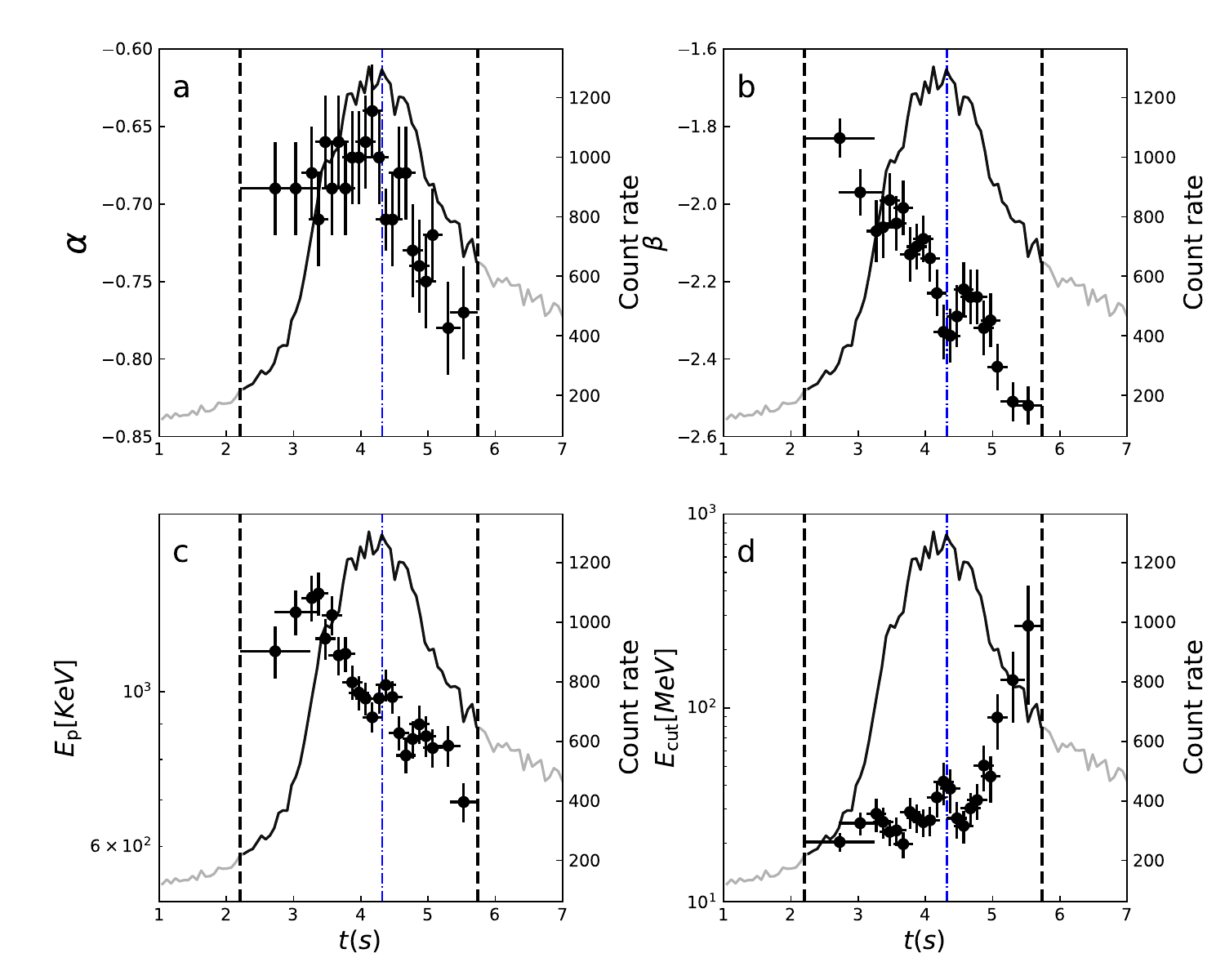}
\caption{Demonstrations of the evolution features of the four BandCut spectral fitting parameters, $\alpha$ (a), $\beta$ (b), $E_{\rm p}$ (c), and $E_{\rm cut}$ (d), which comes from Table 1 in \cite{Lin2019}, and measured in the first pulse of the second emission episode in GRB~160625B. It's corrresponding light curve in 64 ms bins in brightest NaI detecter is over-plotted with solid lines in the right y-axis in each of sub-panel. Here, $t$=$t_{\rm obs}$-$t_0$, where $t_{\rm obs}$ is the time since trigger time ($T_0$), and $t_{0}=185$ s the pulse zero time relative to $T_0$. The two black vertical dashed lines mark the time period (the highlighting solid phases) for our analysis, and the blue vertical dotted dashed lines marks the location of $t_p=4.32$ s , where the flux begins to decay.   }
\label{fig:160625B}
\end{figure}

Shown in Fig.\ref{fig:160625B} are the evolutions of the fitted parameters for the BandCut function in the first pulse of the second episode of GRB~160625B, which comes from Table 1 in \citep{Lin2019}. In this analysis, we set $t_0=185$ s as the offset of the pulse zero time relative to the GRB trigger time ($T_{0}$), which is the time corresponding to one-tenth of the peak flow in the rising phase as done in \cite{Lu2018} (Noted: any values adopted for $t_{0}$ do not affect our analysis results). For a comparison, it’s corrresponding light curve with 64 ms binsize in brightest NaI detecter is also over-plotted in the right y-axis. Thus, the observed features could be summarized as follows: 

(I) The evolution in $E_{\rm p}$ throughout the pulse takes on a so-called intensity-tracking pattern \citep[e.g.][]{Golenetskii1983,Liang1996,Preece2000,Lu2010,Lu2012} , which could be achieved by the numerical simulation in the case that the cooling processes of electrons are dominated by adiabatic cooling \citep{Gao2021}.

(II) The high-energy index, $\beta$, tends to soften progressively throughout the pulse.

(III) For the low-energy index, in the rising phase, it firstly has a less change with a average of $\langle\alpha\rangle=-0.675\pm0.017$ (also see red data points in the left panel of Fig. \ref{fig:alpha-y}), indicating that the underlying radiation physics is the synchrotron emission in the slow-cooling regime \citep[e.g.][]{Sari1998,Ghisellini2000}, then follows by a weak progressive softening trend in the decay phase (further explanation could be found in Section \ref{sec:adiabatic}).

(IV) There is a weak positive correlation between $E_{\rm cut}$ and it's energy flux in the rising phase, following by a significant anticorrelation between them in the decay phase, and a turn-over point rightly takes place at the location where the peak flux begins to decay, as marked by $t_p=4.32$ s relative to $t_0$ in the figure. 

\begin{figure}
\centering
\includegraphics[scale=0.75]{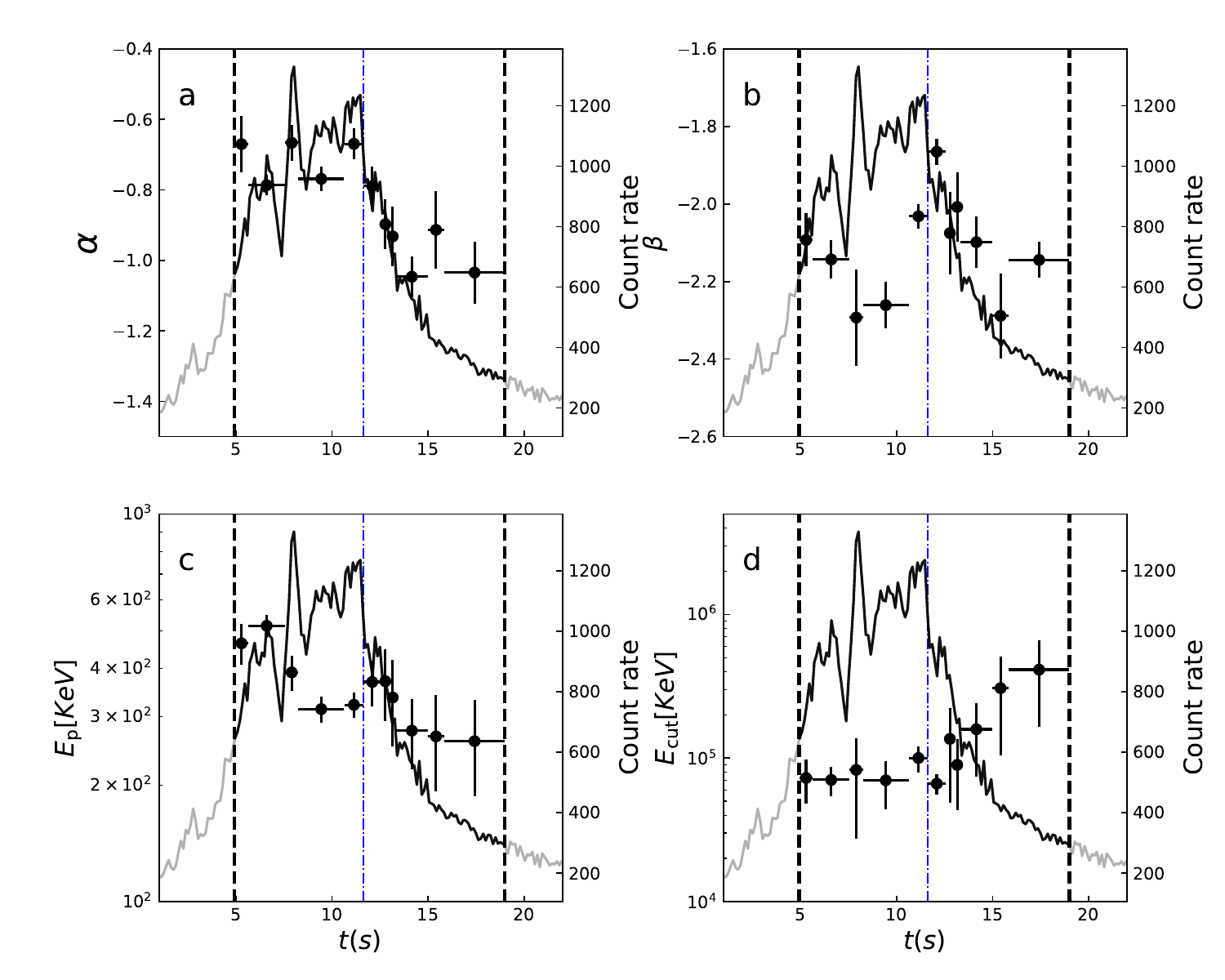}
\caption{The same as Fig.(\ref{fig:160625B}) but for GRB~160509A. Here, $t$=$t_{\rm obs}$-$t_0$, where $t_{0}=6$ s. The blue vertical dotted dashed lines marks the location of $t_p=17.62$ s, where the flux begins to decay.}
\label{fig:160509A}
\end{figure}

Applying the same data analysis method in \cite{Lin2019} to the first main episode in GRB~160509A, we find that the Bandcut function can also provide a good description for their time-resoved spectra. Our analysis results are reported in  Fig. (\ref{fig:160509A}). To demonstrate the consistency of the model with the data, three typical spectral energy distributions given in $F_\nu$ are also shown in Appendix~\ref{Appendix}. In our time-resoved spectral analyses, different from \cite{Vianello2018}, here we apply the Bayesian Blocks algorithm \citep{Scargle2013} to the light curve from the brightest NaI detecter, and also identify the similar spectral evolution features to those in \cite{Vianello2018}. When compared to GRB~160625B, one could find that GRB~160509A shares the same spectral evolution behaviors, except that it is difficult to identify the evolution patterns of $E_{\rm p}$ because of some sub-pulses overlapping in its rising phase. Interestingly, a significant anticorrelation between $E_{\rm cut}$ and it's energy flux is also identified in the smoothly decay phase of GRB~160509A . 

\begin{figure}
\centering
\includegraphics[scale=0.75]{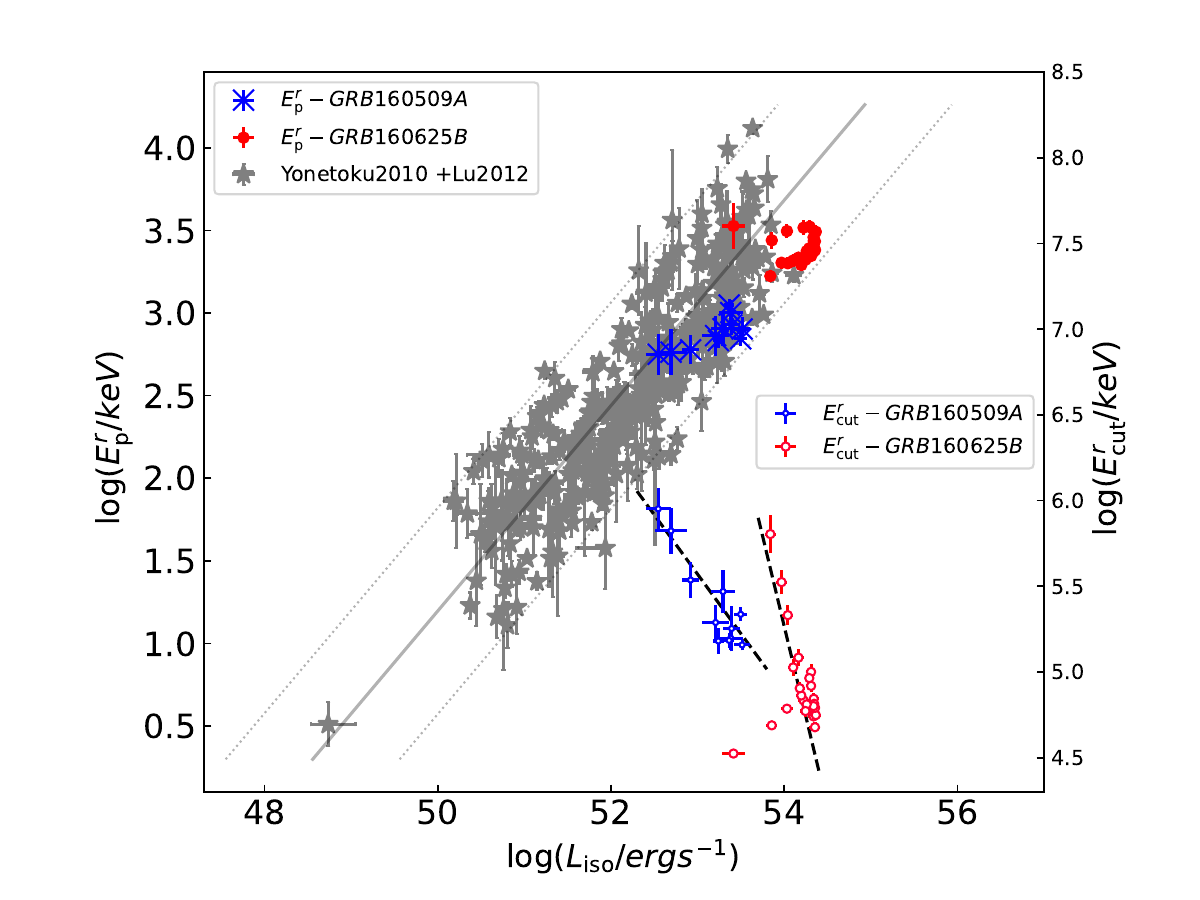}
\caption{Comparison of the time-resolved $E^{r}_{\rm p}-L_{\rm iso}$ correlation (left y-axis) for both GRB~160625B and GRB~160509A to that for other sample (gray `*': the 101 BATSE GRBs in \cite{Yonetoku2010} + the 15 Fermi GRBs in \cite{Lu2012}), The gray solid line is the best fit to the data, while the gray two dotted lines represent its 2$\sigma$ dispersion around the best fit. The time-resolved $E^{r}_{\rm cut}-L_{\rm iso}$ correlation for both GRB~160509A (Noted that the data points both in the rising and decay phases are over-lapped each other) and GRB~160625B (right y-axis) are also plotted in the figure for a comparison. The black dash lines are the best fit correlations: $\log (E^{r}_{\rm cut})=(119.07 \pm 8.57) + (-2.11 \pm 0.16) \log(L_{\rm iso})$  for GRB160625B and $\log (E^{r}_{\rm cut})=(42.28 \pm 4.92) + (-0.69 \pm 0.09) \log(L_{\rm iso})$  for GRB160509A. Noted that the data points in rising phases are excluded from our fittings.}
\label{fig:specenergy}
\end{figure}

Spectral energy correlations for GRBs have been widely studied \citep[e.g.][]{Liang2004,Band2005,Kocevski2012,Nakar2005,Lu2012,Nava2012,Guiriec2015}, their physical origin is still under debate. Here we also identify a positive correlation between $E^{r}_{\rm p}$ and $L_{\rm iso}$ ($E^{r}_{\rm p}=(1+z)E_{\rm p}$, the superscript $r$ denotes the quantities in the rest frame hereafter) in the two bright GRBs in Fig.(\ref{fig:specenergy}). When compared to those from \cite{Lu2012}, in which all spectral analyses are based on the Band function \citep{Band93}, we find that they obey an identical power-law relationship within 2$\sigma$ dispersion. This is not surprising because the high-energy spectral indexes derived from both GRB~160625B and GRB~160509A, ranging from -2.5 to -1.8 (see Figs. of (\ref{fig:160625B}) and (\ref{fig:160509A})), have a similar range to that either from CGRO/BATSE GRB sample \citep{Kaneko2006} or from  Fermi GRB sample \citep{Lu2012}, in which their high-energy spectral indexes are all derived from the Band function. This indicates that they share the same physical origin.

Interestingly, contrary to the $E^{r}_{\rm p}-L_{\rm iso}$ correlation, in both GRB~160625B and GRB~160509A, there exist a nagetive correlation between $E^{r}_{\rm cut}$ and  $L_{\rm iso}$ in their decay phases (hereafter referred to as the  $E^{r}_{\rm cut}-L_{\rm iso}$ relation). Thus we perform a log-linear fit to the data, and obtain the best fitting results: $\log (E^{r}_{\rm cut})=(119.07 \pm 8.57) + (-2.11 \pm 0.16) \log(L_{\rm iso})$  for GRB160625B, and $\log (E^{r}_{\rm cut})=(42.28 \pm 4.92) + (-0.69 \pm 0.09) \log(L_{\rm iso})$  for GRB160509A. The results are also shown in Fig.(\ref{fig:specenergy}) for comparisons.  It is found from the figure that there is no uniform power-law relationship between $E^{r}_{\rm cut}$ and $L_{\rm iso}$ for different GRBs, when compared with the $E^{r}_{\rm p}-L_{\rm iso}$ correlation. Seemingly, the greater the luminosity, the more fast the $E^{r}_{\rm cut}$ decays.

Further, to figure out the temporal evolution of the $E_{\rm cut}$ in the comoving frame, for simplicity, we convert observed quantities into quantities in the comoving frame with the equations: $E^{\prime}_{\rm cut}=E_{\rm cut}(1+z)/(2\Gamma)$ and $t^{\prime}=t_{\rm obs}2\Gamma/(1+z)$ (where $\Gamma$ is the bulk Lorentz factor of the jet ), by taking the time of $t_{p}$ as zero time in the comoving frame, then employ a log-linear fit to the data. The results are shown in Fig.(\ref{fig:Ecut-t}). Intriguingly, it is found that there are almost the same power-law slopes ($\sim$ 1) within their errors\footnote{Noted that the power-law slope is insensitive to the bulk Loentz factor one adopts.}, meaning that $E^{\prime}_{\rm cut}$ increases in time almost with the same behaviors.

\section{Interpretating the observations}\label{sec:adiabatic}
Based on the spectral studies of 11 bursts detected by BATSE, \cite{Cohen1997} for the first time identified the hard low-energy photo spectral indexes, $\alpha \sim -2/3$, in two of them. They further argued that this hard spectra could be generated if the dense region of radiation particles and fields expands rapidly due to internal pressure, and adiabatic losses dominate particle cooling (also see \cite{Ghisellini2000}). Next,  numerical analyses confirm this argument (for more detailed information please refer to \cite{Geng2018,Panaitescu2019,Gao2021}). \cite{Panaitescu2019} also pointed out that, in the context of adiabatic losses dominating the particles cooling, the progressive softening of the $\alpha$ and/or a decrease of the $E_{\rm p}$ are shown to arise naturally from a decreasing magnetic field after electron injection ceases, with some contribution from the spherical curvature of the emitting surface\footnote{According to Equation (47) in \cite{Panaitescu2019} (also see the right upper panel of Figure 5 in their paper), during a pulse decay phase, the weakening of the magnetic field would causes the $E_{\rm p}$ to decay faster than the contribution of the curvature effect.}. Therefore, following the vein, in the next two sub-sections, We try to verify that the observations agree with the model predictions.

\subsection{Case of the adiabatic cooling of the emitting particles} \label{adiabatic}

An electron in a relativistic jet will lose energy when it is moving in the magnetic field with a Lorentz factor $\gamma_{\mathrm{e}}^{\prime}$ \citep{Rybicki1986},
\begin{equation}
\dot{\gamma}_{\text{e,syn}}^{\prime}=-\frac{\sigma_{T} B^{\prime 2} \gamma_{\mathrm{e}}^{\prime 2}}{6 \pi m_{\mathrm{e}} c},
\label{syn}
\end{equation}
where $\sigma_{T}$ is the Thomson cross-section.
Meanwhile, the electron undergoes an adiabatic cooling in the moving jet  \citep{Panaitescu2019,Gao2021},
\begin{equation}
\dot{\gamma}_{\mathrm{e}, \mathrm{adi}}^{\prime}=-\frac{2}{3}\frac{\gamma_{\mathrm{e}}^{\prime}}{t+t_0}=-\frac{2}{3}\frac{c \Gamma \gamma_{\mathrm{e}}^{\prime}}{R}
\label{adi}
\end{equation}
%\begin{equation}
%\dot{\gamma}_{\mathrm{e}, \mathrm{adi}}^{\prime}=-\frac{2}{3}\frac{\gamma_{\mathrm{e}}^{\prime}}{R} \frac{d R}{d t^{\prime}}=
%-\frac{2}{3}\frac{c \gamma_{\mathrm{e}}^{\prime} \Gamma}{R},
%\label{adi}
%\end{equation}
where $R=c(t+t_0)\Gamma$ is the source radius, and $t_0$ is the time since electron injection began. Therefore, the adiabatic cooling will dominate the cooling process when
\begin{equation}
\label{adisyn}
\frac{\dot{\gamma}_{\mathrm{e}, \mathrm{adi}}^{\prime}}{\dot{\gamma}_{\text{e,syn}}^{\prime}}=
\frac{4\pi  m_{\mathrm{e}} c^2 \Gamma}{R \sigma_{T} B^{\prime 2} \gamma_{\mathrm{e}}^{\prime}} \geqslant 1.
\end{equation}
Assuming that the Lorentz factor of the electrons that radiate at the GRB spectral peak energy $E_{\rm p}$ is $\gamma_{\mathrm{e}}^{\prime}$ then we have
\begin{equation}
\label{Ep}
\gamma_{\mathrm{e}}^{\prime}=(\frac{4(1+z)\pi m_e c E_{p}}{3hq_eB^{\prime}\Gamma})^{1/2}.
\end{equation}
Based on Equtions (\ref{adisyn}-\ref{Ep}), one would find that adiabatic cooling will dominate the particle cooling when $B^{\prime}=1$G,$\Gamma=300$, $E_{\rm p}=1 MeV$, and $R<10^{16} $cm.

Based on the comprehensive analyses to adiabatic and radiative cooling of relativistic electrons with an initial power-law electron distribution, $N(\gamma^{\prime}) \sim \gamma^{\prime -p}$, for $\gamma^{\prime}_i < \gamma^{\prime}$, where $\gamma^{\prime}_i$ is minimum injection lorentz factor, \cite{Panaitescu2019} found that a power-law particle injection rate, $k_{i}=t^{\prime y}$, would result in an electron distribution in the cooling-tail, $N(\gamma^{\prime}) \sim \gamma^{\prime -m}$ ( $\gamma^{\prime} < \gamma^{\prime}_i$), with $m=-(3y+1)/2$. This result holds only for $m < p$, whereas, if $-(3y+1)/2 > p$, then $m=p$. Thus the low-energy slope of the photon spectrum $dC/d\varepsilon \sim \varepsilon^{\alpha}$ (The ${\alpha}$ is equal to the low-energy index of the Bandcut function) at energies below the synchrotron characteristic frequency of the $\gamma^{\prime}_i$ electrons  is related to the injection power-law index $y$, as follows (also see  Equation (24) in \cite{Panaitescu2019}) 
\begin{equation}
\alpha=\left\{
\begin{array}{lll}
-\frac{2}{3}  \quad \quad -\frac{5}{9} < y  \quad \quad (m < \frac{1}{3})\\
\frac{3y-1}{4} \quad \quad -\frac{2p+1}{4}<y < -\frac{5}{9} \quad \quad ( \frac{1}{3} < m < p)\\
-\frac{p+1}{2} \quad \quad  y < -\frac{2p+1}{3} \quad \quad (  m = p)
\end{array}.
\right.
\label{cooling}
\end{equation}
For injection power-law index with $y>-5/9$, the cooling-tail is harder than $\gamma^{\prime -1/3}$, and its synchrotron photon spectrum is $~ \varepsilon^{-2/3}$ independent of $y$. Their further investigation revealed that the pulse shape dependences on both $k_{i}$ and the magnetic field $B$,  which are determined by Equations (51)-(54) in their paper. According Equation (\ref{cooling}), the fact that the hard mean low-energy photo spectral indexes, $\langle\alpha\rangle \simeq-2/3$, could be measured in the rising phases of both GRB~160625B and GRB~160509A, would indicate that it is adiabatic cooling process with an increasing injection rate ($y>-5/9$) that yields the pulse rises. As the injection decreases ($y<-5/9$) or is switched off, and the injected electrons migrate toward lower energies, coupling with a decreasing $B$ (also see next sub-section), this cooling process would yield their corresponding pulse decays as the electrons cool below the observing window.  

Fig. (\ref{fig:alpha-y}) shows the compatibility of the mode in the cooling-tail $\gamma^{\prime} < \gamma^{\prime}_{i}$ (Equation (\ref{cooling})) to the observed spectral evolution in the contex of the adiabatic cooling. Based on Equation (\ref{cooling}), we find that the electron injection indexes $y$ derived from the observed low-energy indexes $\alpha$ are all less than the limid $y<-5/9$ (see the pluses `+' in the figure). Meanwhile, the injection indexes satisfies the conditions constrained by the observed high-energy indexes $\beta$ (see the circles in the figure) , i.e., $-\frac{2p+1}{4}<y < -\frac{5}{9}$ (here the relation $p=-(2\beta+1)$ is adopted). This fact shows that the observations in both GRB~160625B and GRB~160509A agree well with the model predictions.

More recently, based on the temporal and spectral features during the tail of bright GRB prompt pulses, \cite{Ronchini2021} found a  novel relation between the spectral index and the flux, then argued that the combined action of the adiabatic cooling of the emitting particles and magnetic field decay can robustly reproduce the relation by using a Bayesian approach and Markov chain Monte Carlo (MCMC) sampling based on the Python package {\tt emcee} \citep{Foreman-Mackey2013}. Inspired by this, in the light curve decay phases of both GRB~160625B and GRB~160509A, we also identify a similar relationship between the high-energy index, $\beta$, and the flux  integrated in the (10-$10^6$) keV band (hereafter referred to as the $\beta-F$ relation).  The results are plotted in Figs. of (\ref{fig:beta-F-160625B}) and (\ref{fig:beta-F-160509A}). 

In order to verify whether the adiabatic cooling model could reproduce the $\beta- F$ relation, we also perform an analogous MCMC analysis as done in \cite{Ronchini2021} based on their public codes\footnote{https://github.com/samueleronchini/Nature\_communications}. The setup of our model parameters is described in the following points (please refer to \cite{Ronchini2021} for details): 

(1) A Bandcut spectral model\footnote{Bandcut: https://asd.gsfc.nasa.gov/Takanori.Sakamoto/personal/},
\begin{equation}
N(E) \propto\left\{
\begin{array}{ll}
(\frac{E}{\mbox{100 keV}})^\alpha \mbox{exp}(-\frac{E}{E_0}), & E < E_{\rm b} \\
(\frac{E}{\mbox{100 keV}})^\beta \mbox{exp}(-\frac{E}{E_{\rm cut}}), & E > E_{\rm b} 
\end{array},
\right.
\end{equation} 
is adopted, where $E_{\rm b}=\frac{E_0 E_{\rm cut}}{E_{\rm cut}-E_0}(\alpha-\beta)$, and $E_{\rm p}=(2+\alpha)E_{\rm b}$. Here, the low energy spectrum index is a constant, i.e., $\alpha=-2/3$ in the case of slow cooling regime by taking $\gamma^{\prime}_c/\gamma^{\prime}_m=30$ for GRB~160625B and $\gamma^{\prime}_c/\gamma^{\prime}_m=20$ for GRB~160509A (motivated by their observed spectral features), respectively. While the high energy index $\beta$ is related to the power-law electron distribution as $\beta=-(p+1)/2$ based on equation (9) in \cite{Ronchini2021}. 

(2) The magnetic field is assumed to evolve as $B^{\prime} \propto (\frac{R}{R_0})^{-\lambda}$ during the adiabatic cooling timescale, defined as $\tau_{ad}=R/2c\Gamma^2$, where $R_0$ is the radius at which adiabatic cooling starts to dominate the evolution of the emitting particles. In this scenario, the gradual fading and softening of the source cause its corresponding spectrum to soften. $\tau_{ad}$ and $\lambda$ are free parameters in the MCMC sampling, respectively. As the model is insensitive to  $\Gamma$, we performed the analysis fixing $\Gamma=300$.

(3) Adiabatic cooling processes take place at a thick shell, i.e., a comoving thickness of the emitting shell $\Delta R^{\prime}=const$.

The  best fit models are also plotted in Figs. (\ref{fig:beta-F-160625B}) and (\ref{fig:beta-F-160509A}) for comparisons. It is found that the combined action of the adiabatic cooling of the emitting particles and magnetic field decay can reproduce the $\beta- F$ relation well. This is further evidence that it is the adiabatic cooling of the emitting particles in a poynting-flux-dominated outflow that powers the two bright GRBs. We obtain an optimal value of $\lambda$ in the range of $1<\lambda<2$, which is expected in a thick emission shell with a softened transverse magnetic field of $B^{\prime} \sim R^{-1}$ in the cases of the jet is conical and adiabatic cooling dominantes the cooling processes \citep{Ronchini2021}.

With the range of $\tau_{ad}$ obtained from our analysis, we could derive their corresponding adiabatic cooling emission radius is $R_{0}=(1.4^{+0.9}_{-1.1})\times 10^{17} (\Gamma/300)^2 $ cm for GRB~160625B and $R_{0}=(1.7^{+1.3}_{-1.1})\times 10^{18} (\Gamma/300)^2 $ cm for GRB~160509A, respectively. Although it is usually expected that the magnetic energy in GRB jet dissipates far from the base of the jet \citep{Lyutikov2003,Zhang2011,Beniamini2016}, these radii are very large and close to the typical deceleration radius of GRB jet, $R_{\rm dec}\sim 10^{17} $cm \citep{Molinari2007,Liang2010}. This may be due to some observational effects\citep{Norris1996,Lu2012,Hakkila2011}, e.g., an unclear onset time of pulse peak and contamination from the superposition of adjacent pulse (see Figs. (\ref{fig:beta-F-160625B}-\ref{fig:beta-F-160509A}). The combination of these effects may lead to bias in the estimation of adiabatic cooling time scales($\tau_{ad}$).

\subsection{Case of Particle Acceleration by Relativistic Magnetic Reconnection }\label{sec:Reconnection}
Inspired by recent particle-in-cell simulations of relativistic turbulence, which show that electrons are impulsively heated in intermittent current sheets by a strong electric field aligned with the local magnetic field,

2D/3D PIC simulations shows that relativistic magnetic reconnection in a poynting-flux-dominated outflow can accelerate particles into a power-law energy distribution with a high-energy exponential cutoff at $\gamma^{\prime}_{cut}$, as follows
\begin{equation}
N(\gamma^{\prime})\propto 
(\frac{\gamma^{\prime}} {\gamma^{\prime}_m})^{-p} \exp(-\frac{\gamma^{\prime}}{\gamma^{\prime}_{cut}}),\quad \quad \gamma^{\prime}_m<\gamma^{\prime} ,
\label{eq:cutoff}
\end{equation}
where $\gamma^{\prime}_m$ minimum injection lorentz factor, and $p$ is the power-law index, which is related to the high energy index $\beta$ of the synchrotron photon number spectrum \citep{Rybicki1979} as follows  
\begin{equation}
	N_{E} \propto E^{\beta} \exp(-(\frac{E}{E_{\rm c}})^{0.5}), \quad \quad \beta=-\frac{p+1}{2},
\end{equation}
where $E_{\rm c}=2\Gamma E^{\prime}_{\rm cut}/(1+z)$ and  $E^{\prime}_{\rm cut} \propto B^{\prime}\gamma^{\prime 2}_{cut}$, where $\Gamma$ and $B^{\prime}$ are the bulk Loentz factor of a jet and the magnetic field strength in the comoving frame, respectively. Following the vein of Section \ref{adiabatic} and assuming that the  magnetic field strength in a poynting-flux-dominated outflow decays by the law of $B^{\prime} \propto (\frac{R}{R_0})^{-\lambda}$, coupling with $R \sim c t \propto t$, we get 
\begin{equation}
E^{\prime}_{cut} \propto B^{\prime} \gamma^{\prime 2}_{cut} \propto t^{\prime \xi} , \quad \quad \xi=2-\lambda.
\label{Ecut}
\end{equation}
   
By using the value of $\lambda$ derived from the $\beta-F$ relation, we obtain $\xi=0.74\pm0.08$ for GRB~160625B and $\xi=0.90\pm0.05$ for GRB~160509A, respectively. These values are consistent with that of the 3D PIC simulation\citep{Zhang2021,Werner2021}.

\section{Discussions and Conclutions}\label{sec:diss}

Two pieces of evidence from the observations of two bright GRBs point toward the same picture that high-energy spectral cutoff grows approximately linearly with time, i.e., $E^{\prime}_{\rm cut} \propto t^{\prime \xi}$ with $\xi\sim1$, which is consistent with the simulation result obtained from 3D PIC model \citep{Zhang2021,Werner2021}, indicating that it is relativistic magnetic reconnection that  accelerates high-energy particles so efficiently that its corresponding high-energy spectral cutoff grows linearly with time. Meanwhile, it is the adiabatic cooling  process of the emitting particles in the magnetic reconnection regions that produces the observed $\beta- F$ relationship. This is understandable because in the context of a strongly magnetized outflow, the adiabatic condition is easily satisfied as the gyroresonance scattering is absent with the particle Larmor radius below magnetic turbulence scale\citep{Xu2017}. 

The Yonetoku relation \citep{Yonetoku2010} is usually used to discriminate its underlying radiation mechanisms. In general, spectral properties (such as $E_{p}$ or $E_{\rm cut}$) carry the key to understanding the physics of GRBs, such as the energy dissipation mechanism, the radiation mechanism, jet structure, as well as the properties of the central engine. Observationally, two evolution patterns of $E_{\rm p }$, i.e., hard-to-soft evolution and intensity-tracking \citep{Bhat1994,Golenetskii1983,Kargatis1994,Ford1995,Liang1996,Norris1986,Preece1998,Kaneko2006,Lu2010,Lu2012} have been identified in different bursts, even in a same GRB in different pulses \citep{Lu2012}.  \cite{Zhang2011} argued that a hard-to-soft evolution of $E_{\rm p }$ within a pulse is expected in the case of the sudden discharge of magnetic energy through turbulent magnetic reconnection. More recently, with numerical analysis in the framework of synchrotron radiation, \cite{Gao2021} found that the two evolution patterns of $E_{p}$ could be reproduced well (also see \cite{Deng2014,Uhm2016}), specifically, the dominant adiabatic cooling process of electrons, accompanied by the decay of $B^{\prime}$,  could generate an intensity-tracking pattern in $E_{\rm p }$, while hard-to-soft patterns of $E_{\rm p }$ are normally expected because of the fact that the adiabatic cooling rate is inversely proportional to the radius of the shell \citep{Deng2014}. These theoretical analyses, together with the observational evidence from the two bright GRBs further confirm that synchrotron radiation may be responsible for the $E^{r}_{\rm p}-L_{\rm iso}$ relation.

\begin{figure}
\centering
\includegraphics[scale=0.44]{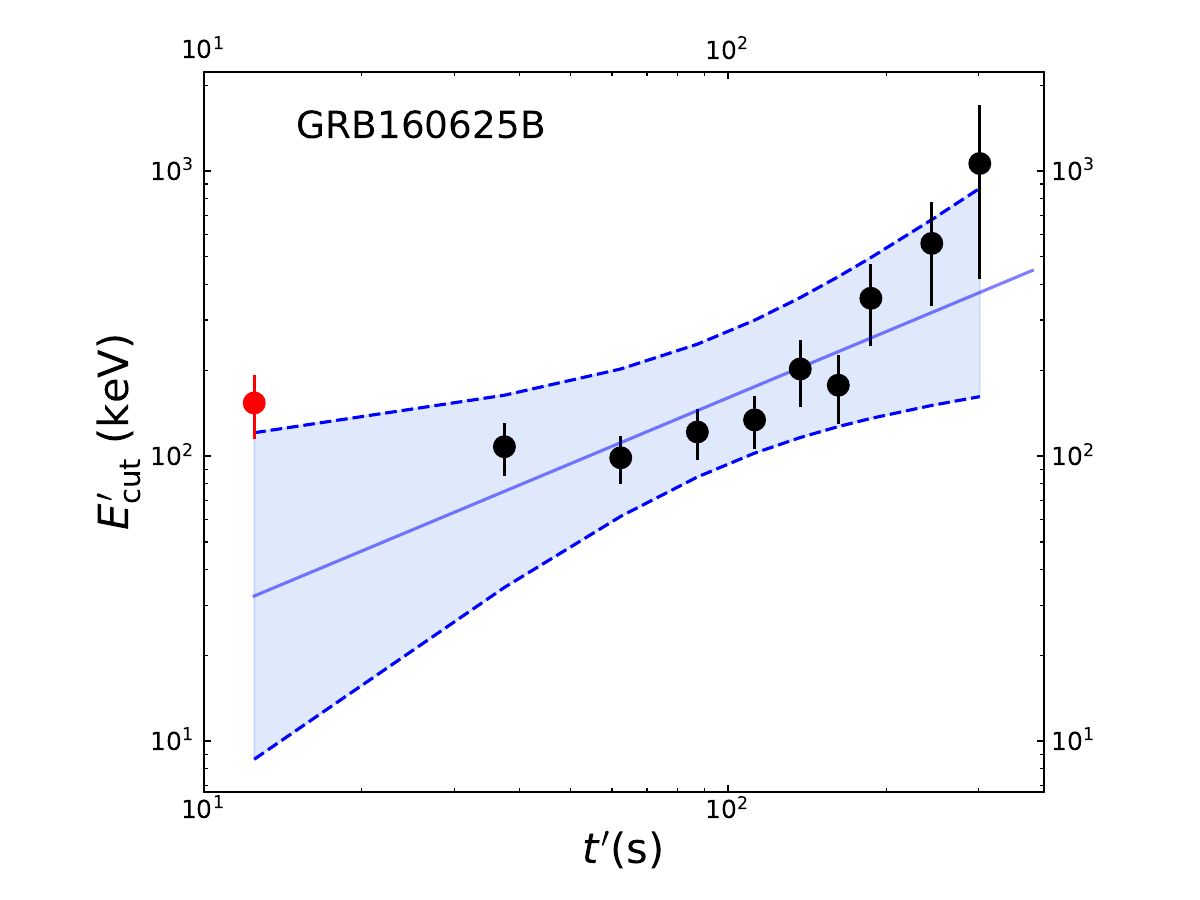}
\includegraphics[scale=0.44]{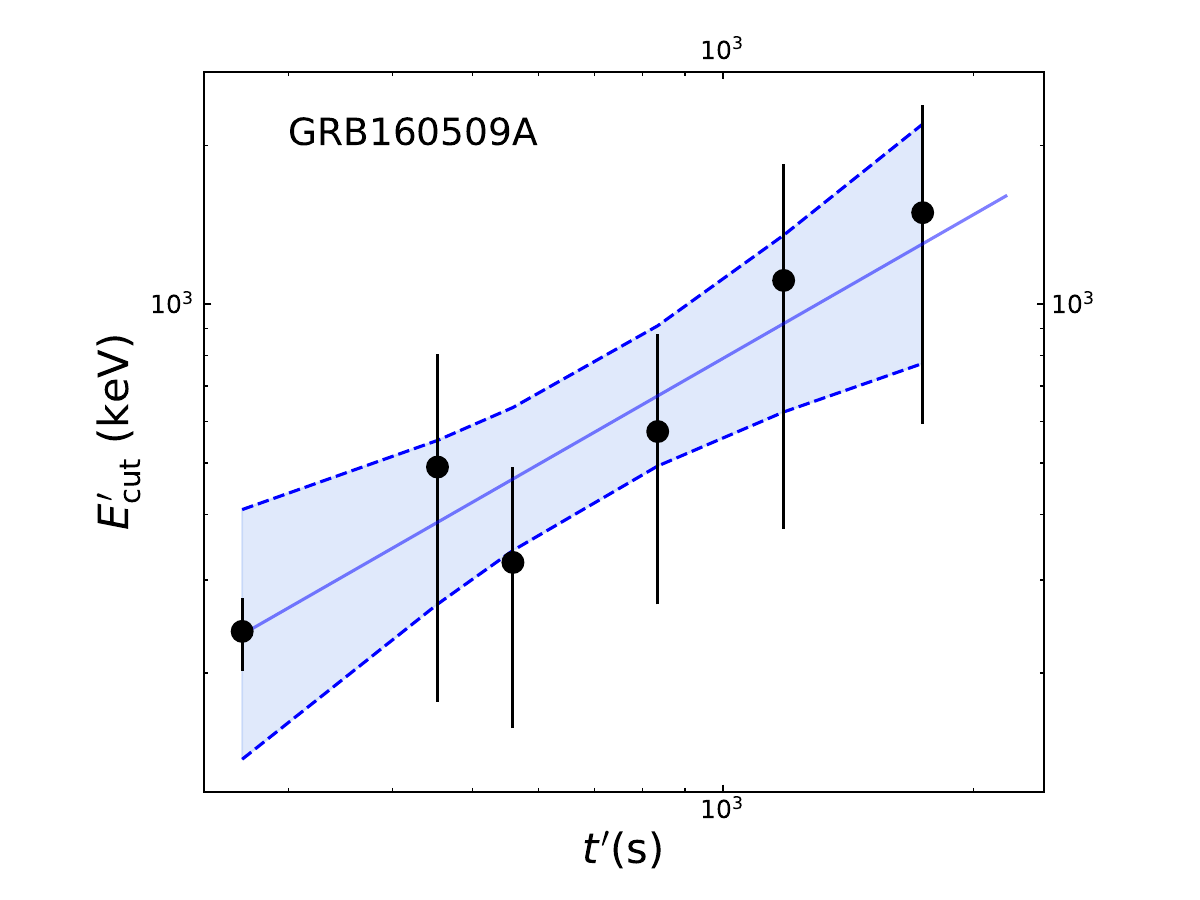}
\caption{Demonstrations of the evolution of $E^{\prime}_{\rm cut}$ in the decay phases of both GRB~160625B (left panel) and GRB~160509A (right panel). Here we take the peak time of $t_{p}$ at which the flux begins to decay as zero time in the comoving frame. The bulk Loentz factor $\Gamma=300$ is adopted for the two GRBs.  The blue solid lines are the best fit correlations: $\log E^{\prime}_{\rm cut}=(0.665 \pm 0.39 )+(0.77 \pm 0.21 ) \log t^{\prime}$ for GRB~160625B  and $\log E^{\prime}_{\rm cut}=(-0.82 \pm 0.34 )+(0.91 \pm 0.24 ) \log t^{\prime}$ for GRB~160509A. The two blue dash lines represent their corresponding 95\% confidence bands. Noted that the red point in the left panel is excluded from the fitting.}
\label{fig:Ecut-t}
\end{figure}

\begin{figure}
\centering
\includegraphics[scale=0.44]{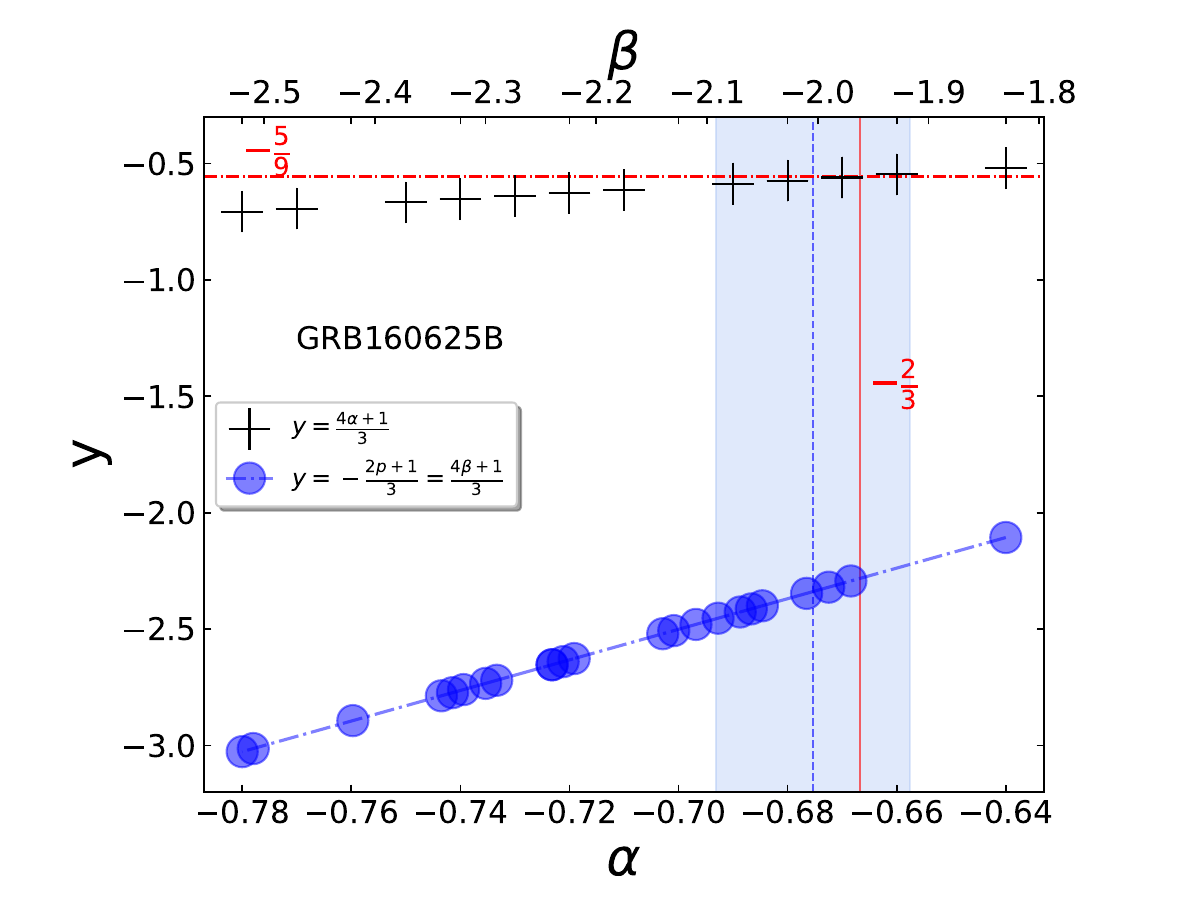}
\includegraphics[scale=0.44]{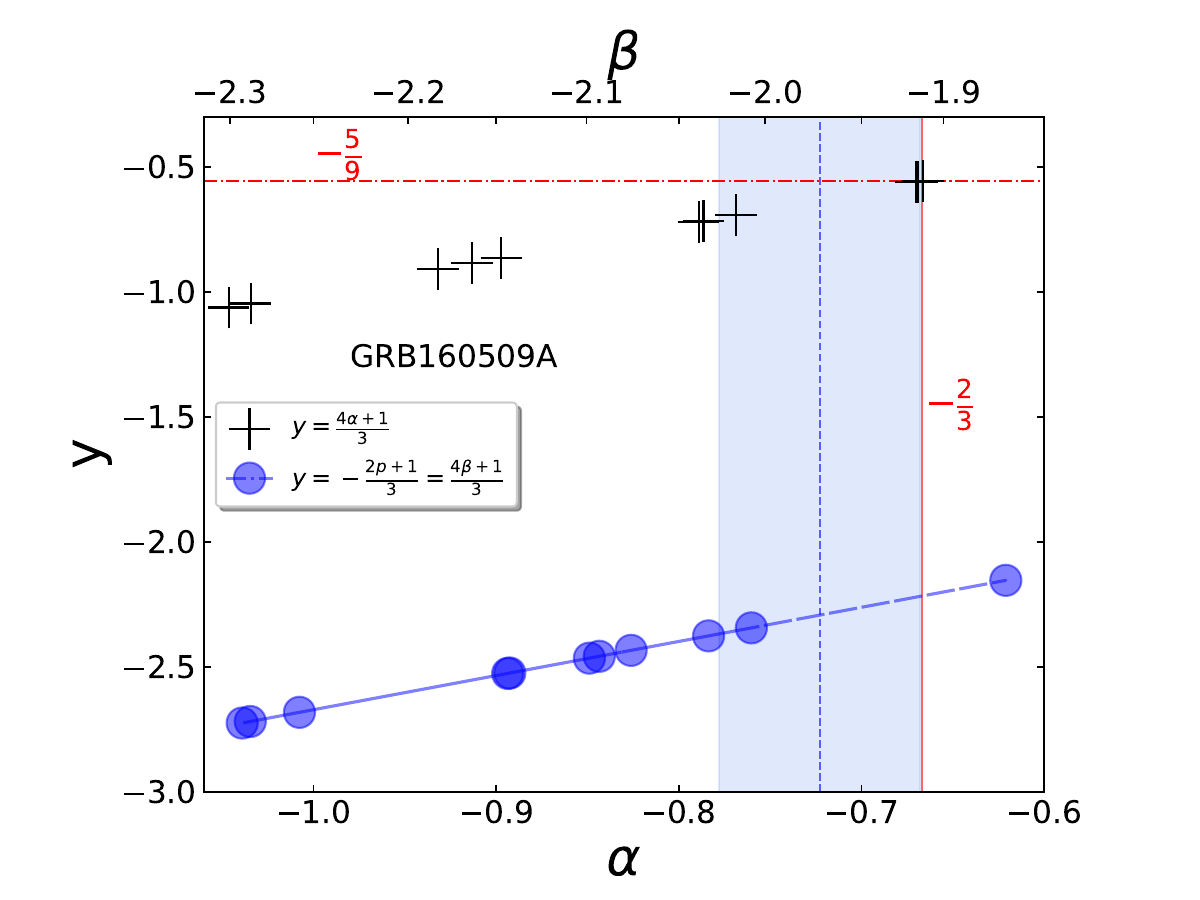}
\caption{Particle injection rate $y$ is related to the low-energy slope $\alpha$ (pluses `+', bottom x-axises) of the photon spectra measured in both GRB~160625B (left panel) and GRB~160509A (right panel) by $y=(4\alpha+1)/3$, whose conditions constrained by the observed high-energy slope $\beta$ i.e., $-\frac{2p+1}{4}< y < -\frac{5}{9}$ are also shown with the circles (top x-axises). The red vertical solid lines and the red horizontal dash-doted lines represent the locations of $\alpha=-2/3$ and $y=-5/9$, respectively. The blue vertical dashed lines and the blue areas represents the average and its 1 $\sigma$ error of the $\alpha$ measured in the rising phases, respectively. Obviously, the blue circles are all located below the pluses, meaning that all the data points satisfy the conditions in Equation (\ref{cooling}).}
\label{fig:alpha-y}
\end{figure}

\begin{figure}
\centering
\includegraphics[scale=0.35]{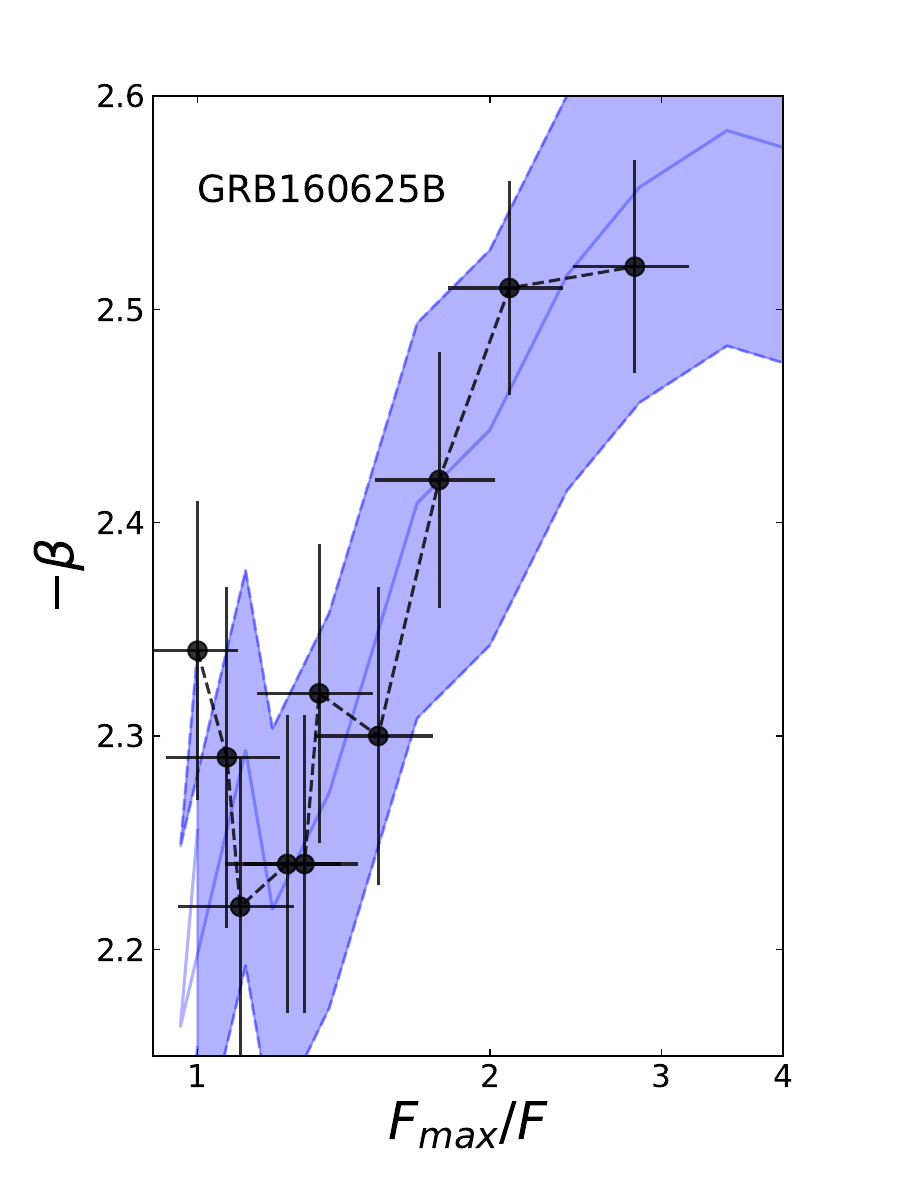}
\includegraphics[scale=0.35]{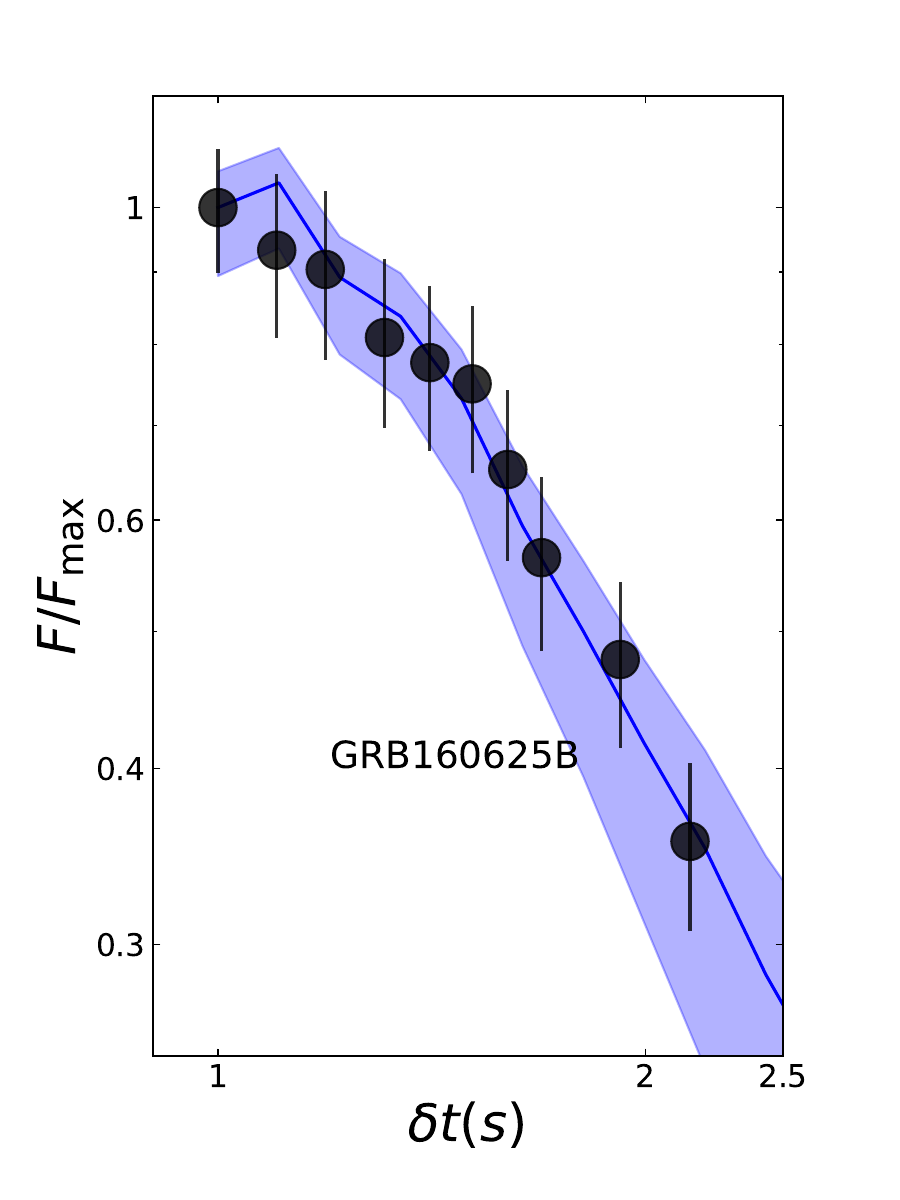}
\includegraphics[scale=0.35]{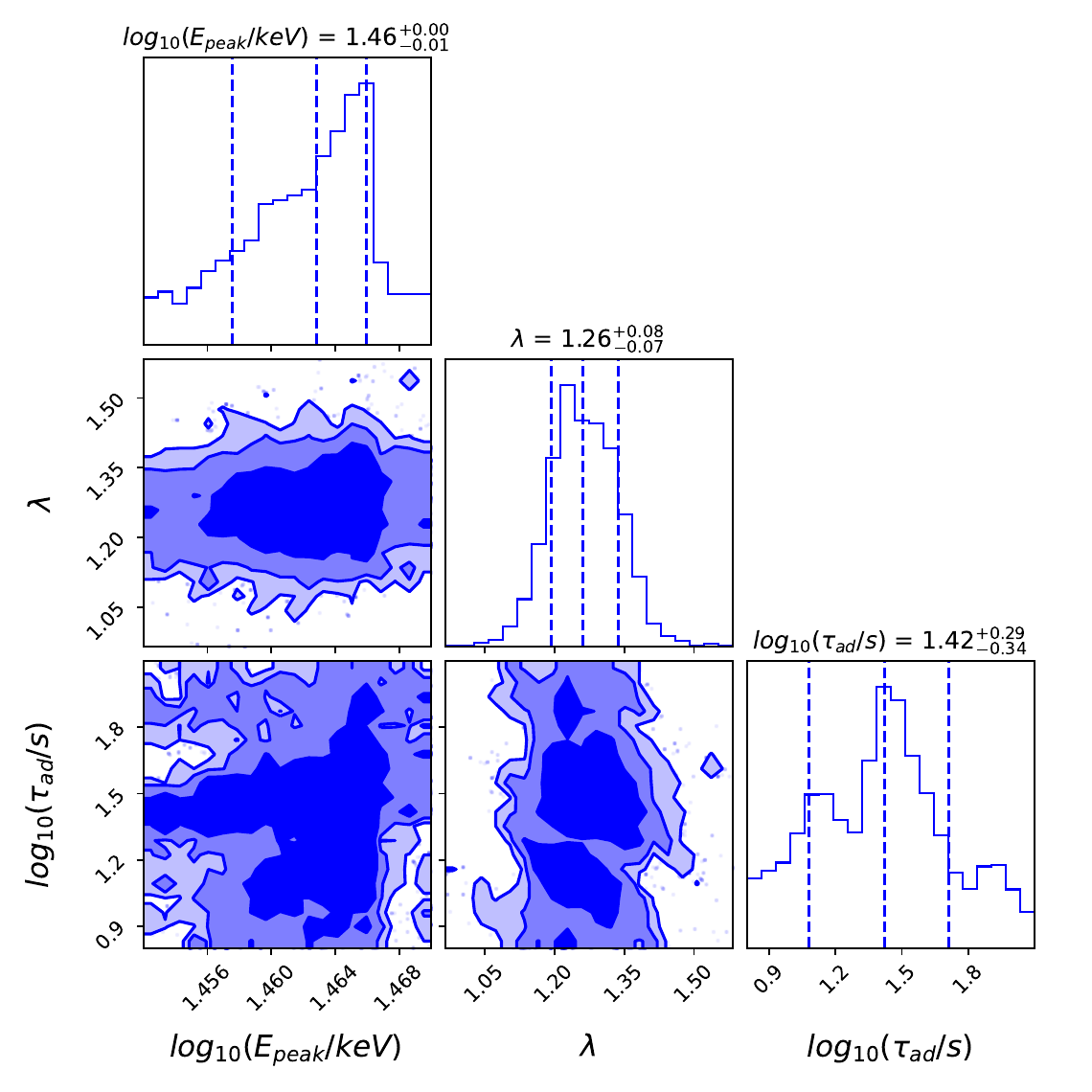}
\caption{Demonstrations of the evolutions of both the hight energy spectral index $\beta$ (left) and the normalized flux $F_{\rm max}/F$ (middle, here $\delta t=t_{\rm obs}-t_{\rm p}+1$) in GRB~160625B, respectively. The blue solid lines are the best fit to the data with the best model parameterrs as shown in the right panel, and the blue areas mark 95\% confidence band of the best fit. Right: A Bayesian approach and MCMC sampling are adopted for the model parameter estimation based on {\tt emcee} package, and the corresponding credible intervals of the model parameters are plotted using the {\tt corner.py} module \citep{Foreman-Mackey2016}, in which 68\%, 90\% and 95\% credible intervals of parameters are also plotted in different colors. Here $E_{\rm peak}$, $\lambda$, and  $\tau_{\rm ad}$ represent the peak energy at $t_{\rm p}$, the decaying index of magnetic field, and the adiabatic timescale, respectively.}        
\label{fig:beta-F-160625B}
\end{figure}

\begin{figure}
\centering
\includegraphics[scale=0.35]{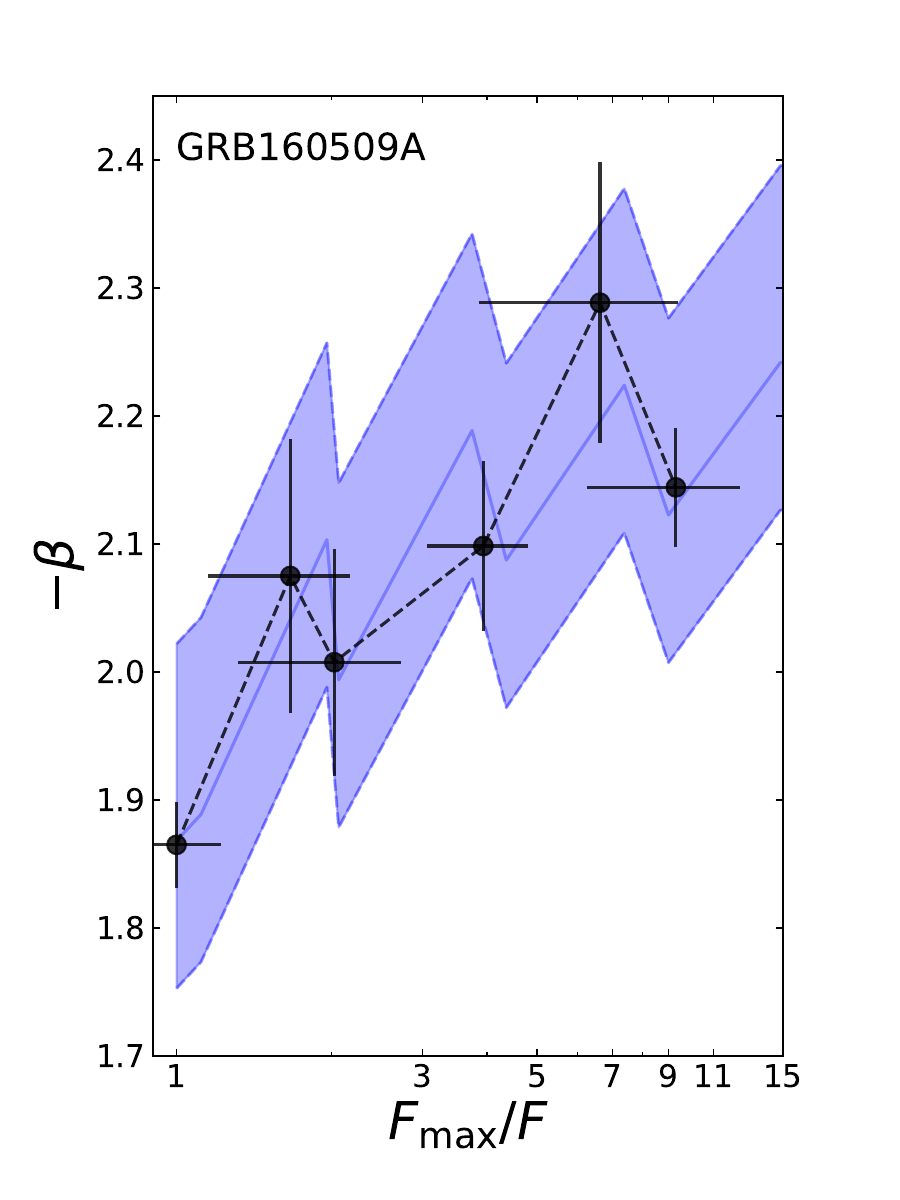}
\includegraphics[scale=0.35]{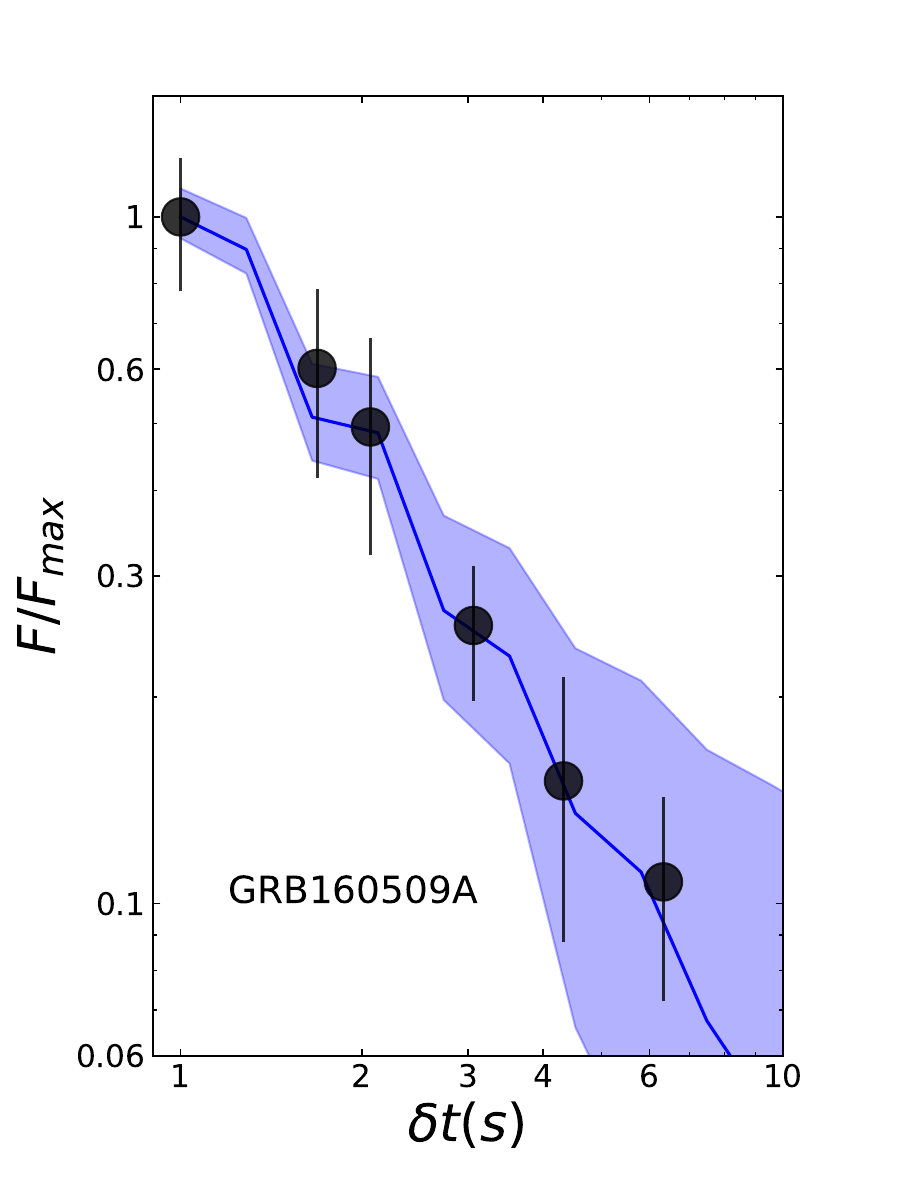}
\includegraphics[scale=0.35]{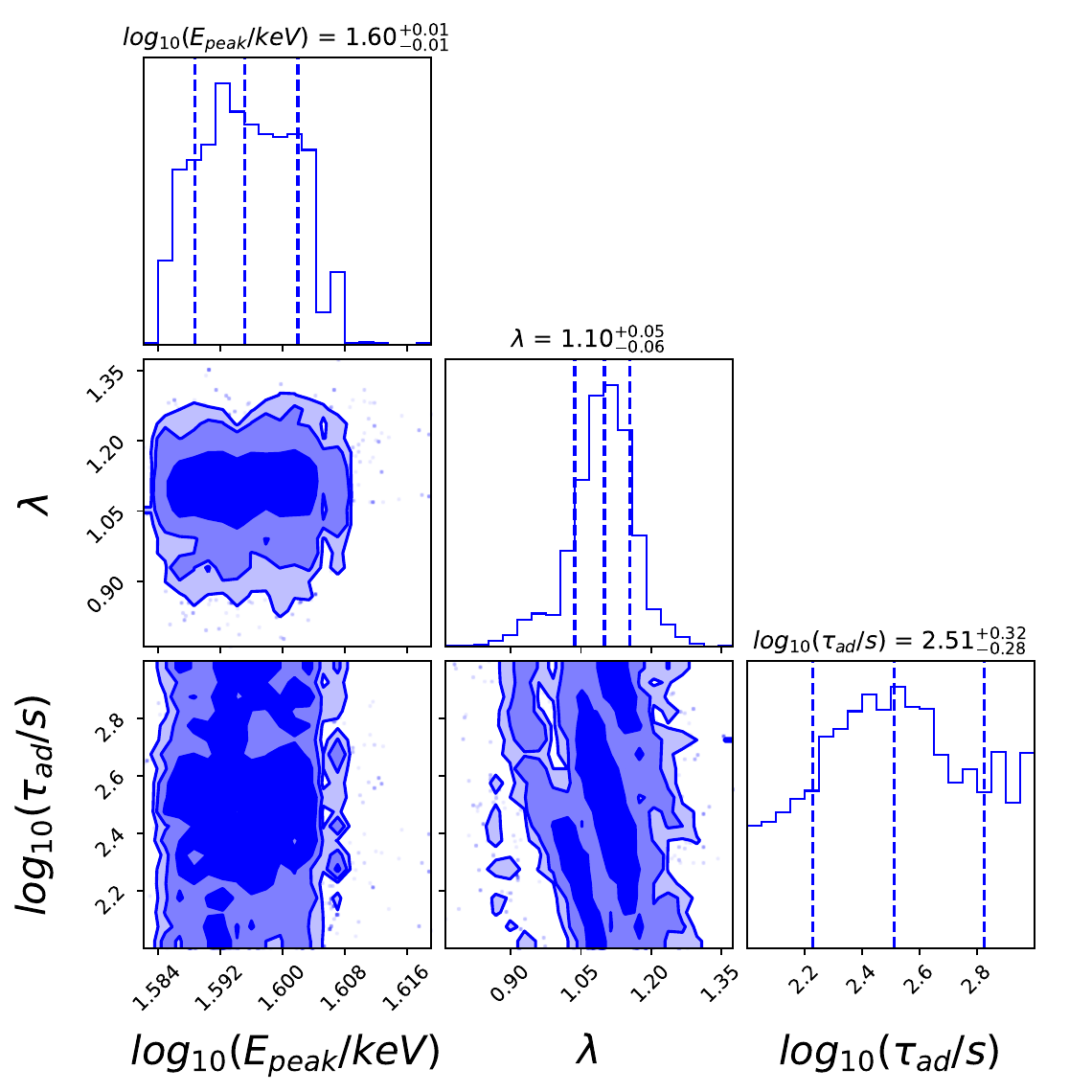}
\caption{The same as Figure \ref{fig:beta-F-160625B} but for GRB~160509A.}
\label{fig:beta-F-160509A}
\end{figure}

At the same time, as mentioned above, in the decay phases of both GRB~160625B and GRB~160509A, we identify a significant anticorrelation between $E^{\prime}_{\rm cut}$ and  $L_{\rm iso}$. Based on the above analyses of the two bright GRBs, we believe that, it is the relativistic magnetic reconnection in a poynting-flux-dominated outflow that accelerates particles into a power-law energy distribution with a high-energy exponential cutoff. As the outflow expands and its magnetic field decreases, the adiabatic cooling of this power-law electron distribution powers their pulse decays. In this case, this anticorrelation is naturally expected due to the following reasons. Firstly, 3D PIC simulations \citep{Zhang2021} show that, compared with most of the low-energy particles located in the downstream region, only a fraction of high-energy particles with $\gamma^{\prime}> 3\sigma$(here, $\sigma$ is the magnetization) can escape from reconnection plasmoids by moving  along the z-direction of the electric current and rapidly accelerated by the large-scale upstream fields. This allows the high-energy spectral cutoff to grow with time as $E^{\prime}_{\rm cut}\propto t^{\prime}$, which is confirmed by the observations from the two bright GRBs (see Fig.\ref{fig:Ecut-t}). Secondly, PIC simulations also found \citep{Werner2021,Li2023L} that the power law spectrum of the particles trapped in plasmoids continuously softens with a decreasing $\sigma$, although the high-energy spectral cutoff sustainably increases with system size. Similarly, \cite{Ronchini2021} argued that a gradual decrease of both the magnetic field and particle injection rate can produce a softening of the spectrum (also see \cite{Panaitescu2019}). Thus, the adiabatic cooling of the softening spectra would cause a drop in luminosity. 

In all, the combination of the two spectral energy correlations, if available, can provide a robust diagnosis of their underlying mechanisms of particle acceleration and radiation. Furtherly, an anticorrelation between $E^{\prime}_{\rm cut}$ and  $L_{\rm iso}$ may be crucial indicator with which to discriminate the particles acceleration by relativistic magnetic reconnection in a variety of astrophysical sources. 

%---------------------------------------------------------------------------------------------------------------------

\clearpage
\begin{acknowledgments}
This work is supported by the National Natural Science Foundation of China (grant No. 12133003). This work is also supported by the Guangxi Talent Program (``Highland of Innovation Talents")

\end{acknowledgments}

%\section*{Code availability}
%Codes used to produce the plots in this paper are available in this public repository:\\
%\url{https://github.com/samueleronchini/Nature\_communications}\\
%XSPEC and PyXspec are freely available online at the following links:\\
%\url{https://heasarc.gsfc.nasa.gov/xanadu/xspec/}\\
%\url{https://heasarc.gsfc.nasa.gov/docs/xanadu/xspec/python/html/index.html}

%------------------------------------------------

\clearpage

%------------------------------------------------

\clearpage



%\bibliography{cite}{}
%\bibliographystyle{aasjournal}

%A idea for the increasing of Ec in the pulse rising phase, because we assume the photon from synchrotron radiation,the synchrotron radiation spectrum have a critical frequency, it is proportional to electron lorentz factor, so when energy injection, the Ec increasing caused by electron lorentz factor increasing.

\clearpage
\appendix

\section{Joint Spectral Fitting}\label{Appendix}
\clearpage
\begin{figure*}
\centering
%\begin{tabular}{ccc}
\includegraphics[scale=0.35]{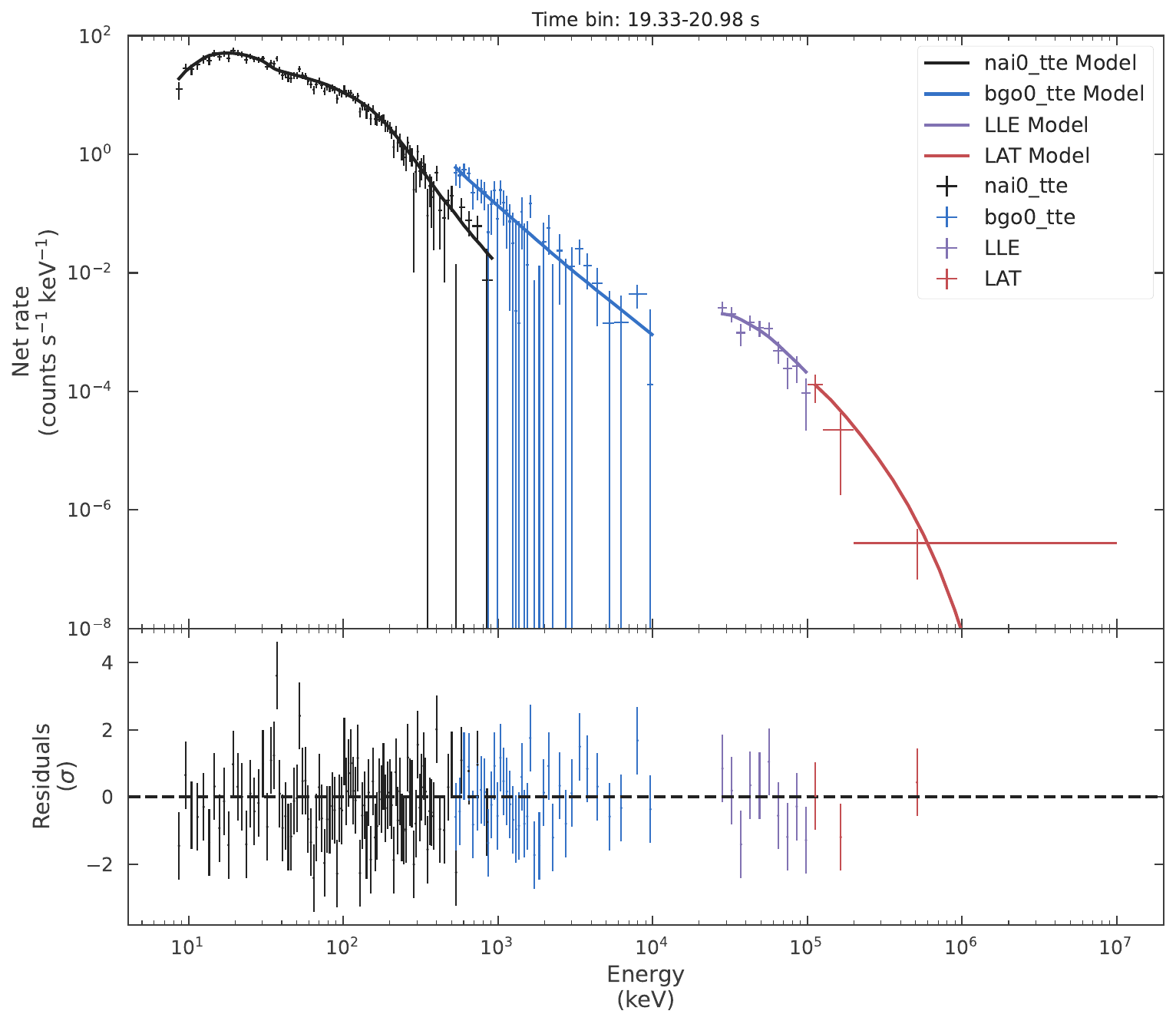}
\includegraphics[scale=0.35]{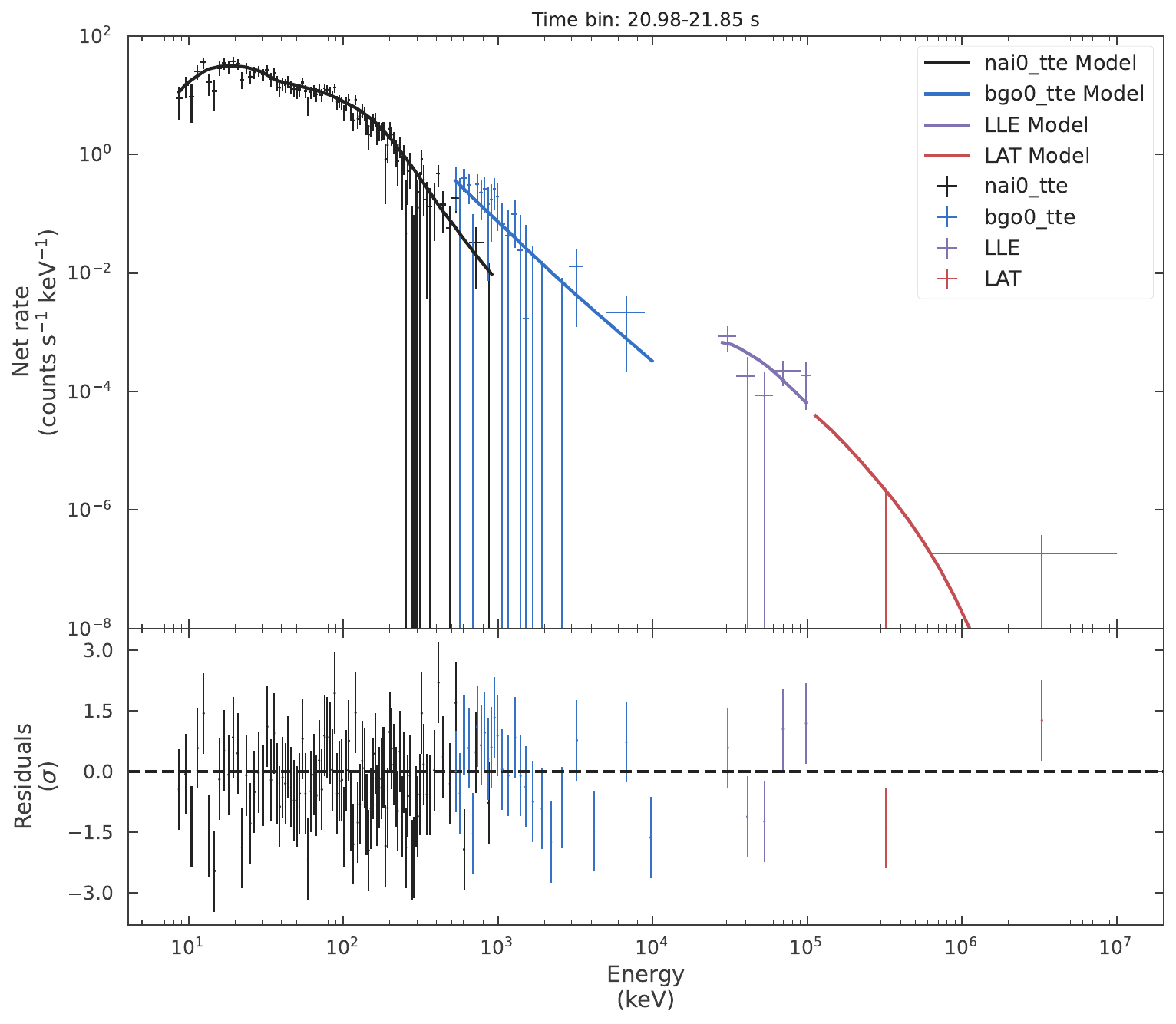}
\includegraphics[scale=0.35]{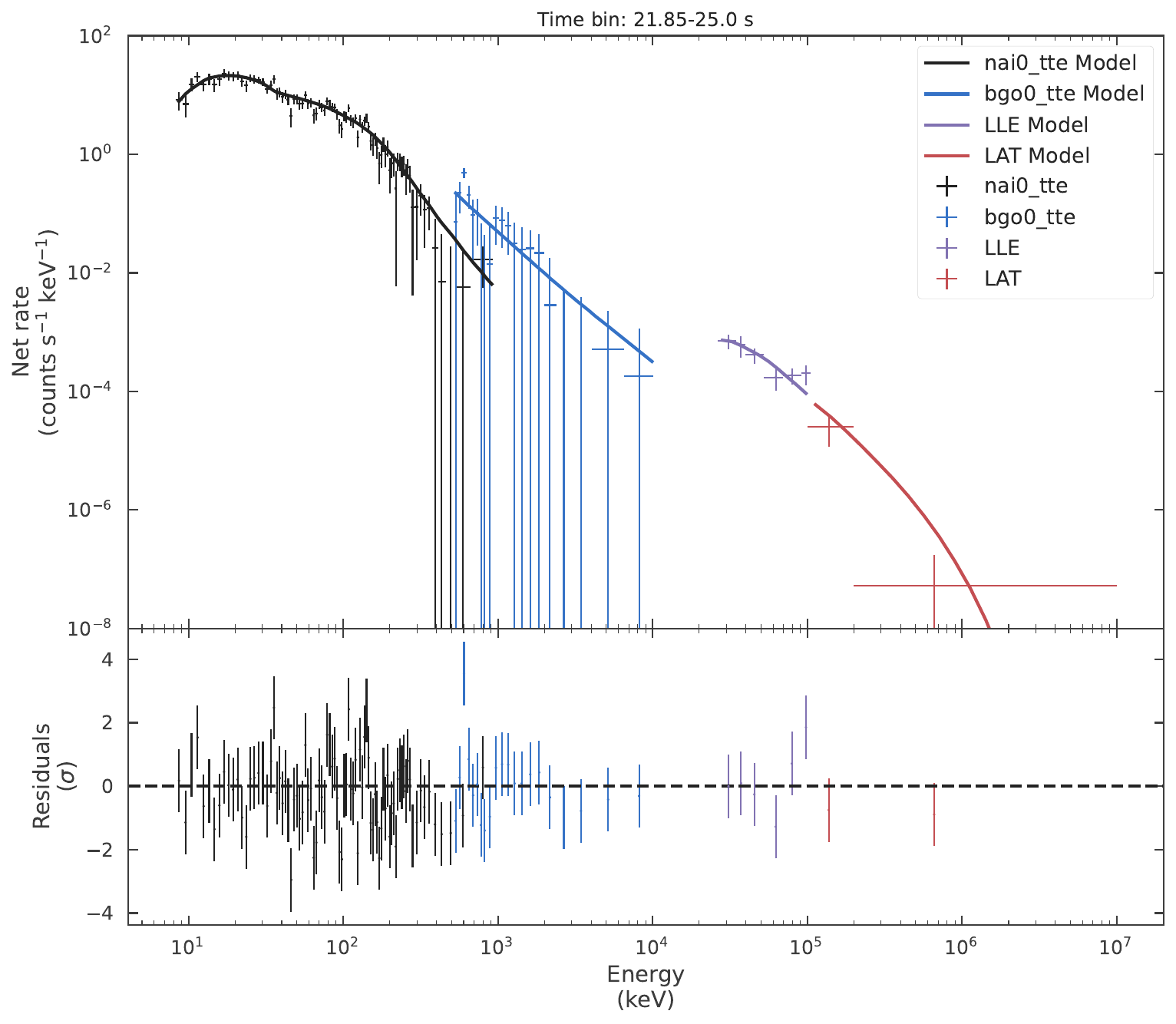}
%\end{tabular}
\caption{Representative spectral fits to GRB~160509A. The details of data reduction and joint spectral fitting could be referred to \cite{Lin2019}.
}\label{Fig:Joint_Spectral_fittings_LAT}
\end{figure*}

\end{document}